\documentclass[aps,prl,reprint,showpacs]{revtex4-1}

\usepackage{amsmath}
\usepackage{amsfonts}
\usepackage{graphicx}
\usepackage{multirow}
\usepackage{subfigure}
\usepackage{color}
\DeclareMathOperator{\Tr}{Tr}

\begin{document}
\title{Experimental Quantum Hamiltonian Identification from Measurement Time Traces}
\author{Shi-Yao Hou}
\author{Hang Li}

\author{Gui-Lu Long}
\email{gllong@tsinghua.edu.cn}
\affiliation{$^1$State Key Laboratory of Low-dimensional Quantum Physics and Department of Physics, Tsinghua University, Beijing 100084, China}
\affiliation{$^2$The Innovative Center of Quantum Matter, Beijing 100084, China}
\affiliation{$^3$Tsinghua National Laboratory of Information Science and Technology,  Beijing 100084, China}

\begin{abstract}
Identifying Hamiltonian of a quantum system is of vital importance for quantum information processing. In this Letter, we realized and benchmarked a quantum Hamiltonian identification algorithm recently proposed [Phys. Rev. Lett. \textbf{113}, 080401 (2014)]. we realized the algorithm on liquid nuclear magnetic resonance quantum information processor using two different working media with different forms of Hamiltonian. Our experiment realized the quantum identification algorithm based on free induction decay signals.  We also showed  how to process data obtained in practical experiment. We studied the influence of decoherence by numerical simulations. Our experiments and simulations demonstrate that the algorithm is effective and robust.
\end{abstract}
\maketitle

\textit{Introduction.}-One critical task is to characterize a quantum system so that it can be used for quantum information processing tasks, such as quantum teleportation \cite{teleportation}, quantum cryptography \cite{BB84,QSDC}, quantum computation \cite{computation1,computation2} and quantum metrology \cite{qmetr}.  One way of fully characterizing a quantum system is doing quantum processing tomography (QPT). The QPT approach requires an exponential number of experiments, which makes it difficult to be realized for even a small sized quantum system \cite{QPT1,QPT2,QPT3,holonmr}.

For general quantum systems, various methods based on measurement time traces for  Hamiltonian identification are proposed. Fourier transformation (FT) of  only one measurement observable is used for a single qubit Hamiltonian identification \cite{sinquid}. Temporal evolution of concurrence measure of entanglement is employed to identify arbitrary two-qubit Hamiltonian \cite{twoquid}. Schemes of estimating the coupling parameters of a complex quantum network based on measurements on a small part of the network is proposed \cite{coupest,inhamid}. A basic and general  framework for quantum system identification on how much knowledge about the quantum system is attainable in principle for a given experimental setup is established \cite{qsi}. Recently, Zhang and Sarovar proposed an approach (ZS approach) for identifying arbitrary Hamiltonian quantum dynamics that takes advantage of available prior knowledge of the system \cite{qhi}.

One typical quantum system is the nuclear magnetic resonance (NMR) system, which is well described by quantum mechanics. Moreover, control technology has been well developed during the 50 years since the birth NMR. These factors make the NMR system an appealing quantum system for sophisticated manipulation. Therefore NMR systems are widely used for quantum information processing \cite{revnmr,nmrcon}. To obtain the information of an NMR system, modern NMR spectrometers acquire the free induction decay (FID) signals, which are the measurement time traces for certain observables. Schemes based on FT (e.g. FT-NMR) of the FID signals, which is one of the most robust way of processing FID,  are developed  \cite{Ernst,spindyn}.  Since ZS approach is also based on measurement time traces and can be applied to arbitrary quantum system,  the NMR spectrometer provides a practical and well controlled system for benchmarking ZS approach.

In this Letter, we experimentally demonstrated the ZS algorithm in NMR. We first perform the ZS approach in an NMR quantum information processor and compared the result with that of FT. The experiment is performed with two kinds of work media with different Hamiltonian forms, due to which, we can show the different experimental setups. Unlike NMR quantum computing experiment, the ZS algorithm starts from mixed state and directly processes the FID signals. Numerical simulations are also performed to analyse the effects of imperfect experiment conditions and decoherence.

\textit{Algorithm.}-Here we briefly introduce ZS approach \footnote{Detailed information are presented in the Supplementrary Material and also can be seen in Ref. \cite{qhi}, which also includes the details of the experiment and numerical simulations.}.  For an $n-$qubit system, all Hermitian operators could be decomposed into the summation of the tensor product of $2\times 2$ identity matrix $I_2$ and Pauli matrices, thus we can choose a set of basis \begin{equation}
S=\{\hat{X}_k|\hat{X}_k=\sigma_{\alpha}^1\otimes\sigma_{\beta}^2\otimes\cdots\otimes \sigma_{\gamma}^n\},
\end{equation}
where $\sigma_{\alpha}$, $\sigma_{\beta}$ and $\sigma_{\gamma}$ are the Pauli matrices $\sigma_x$, $\sigma_y$ and $\sigma_z$ and $I_2$. The number of the elements in $S$ is $4^n$. Then we can decompose the Hamiltonian as\begin{equation}
\label{Hgen}
\hat{H}=\sum_{m=1}^{M} a_m X_m,
\end{equation}
with $X_m \in S$ and $a_m$ being the corresponding parameter.
Then one needs to obtain the dynamics of observables and simplify the dynamics according to the appropriately chosen observable. Since all the elements in $S$ are Hermitian operators, they could be considered as observables. The expectation value $x_k$ for observable $\hat{X}_k$ in state $\rho$, written as $x_k=\Tr\{\rho\hat{X}_k\}$, its time derivative is
\begin{equation}
\dot{x}_k=\sum_{l=1}^{4^n}\left(\sum_{m=1}^{M} C_{mkl} a_m\right)x_l,
\end{equation}
where $C_{mkl}=\Tr\{i\hat{X}_l[X_m,\hat{X}_k]\}$. Typically, for a practical quantum system, only the expectation values of certain observables are accessible in experiments.  Suppose the observable  we can access in experiment is $O=\sum_j o_j \hat{X}_j$. Collect all the $\hat{X}_j$'s presented in the expansion of $O$ in the set  $\mathcal{M}=\{ M_{\nu_1}, M_{\nu_2}, ...,M_{\nu_p} \}$, with $\boldsymbol{\nu}=\left[\nu_1, \nu_2,...,\nu_p \right]^\text{T}$ being a vector of $p$ components.  Let $\Delta=\{X_m\}_{m=1}^M$, and define an iterative procedure as  \begin{equation}\label{gbar}
G_0=\mathcal{M},G_i=\left[G_{i-1},\Delta\right]\bigcup G_{i-1},
\end{equation}
where $[G_{i-1},\Delta]\equiv\{\hat{X}_j:\Tr(X_j^\dagger[g,h])\neq 0,$ where $g \in G_{i-1},$ and $h\in \Delta \},$ finally we will reach at a maximal set $\bar{G}$ after finite steps, which is called accessible set. Writing all the $x_k$ with $\hat{X}_k\in\bar{G}$ in a vector $\mathbf{x}_a$, the dynamics of this vector is \begin{equation}\label{A2}
\mathbf{\dot{x}_a}=\tilde{\mathbf{A}}\mathbf{x}_a,
\end{equation}
where $\tilde{\mathbf{A}}$ is a $K\times K$ matrix with $K$ being the number of elements in $\bar{G}$.  For a time-independent Hamiltonian, Eq. (\ref{A2}) can be solved as ${\mathbf{x}}_a=\exp(\mathbf{\tilde{A}}t)\mathbf{x}_a(0)$. In experiments, data points are obtained with time intervals $\Delta t$. With $\mathbf{x}_a(j)$ denoting $\mathbf{x}_a(j\Delta t)$, we have $\mathbf{x}_a (j) = \exp(\mathbf{\tilde{A}}j\Delta t)\mathbf{x}_a (0)$. Let $y$ denote the expectation value of observable $O$. We can find that $y=\mathbf{C}\mathbf{x}_a$. With $\mathbf{A}_d=\exp(\mathbf{\tilde{A}}\Delta t)$, we have \begin{equation}
y(j)=\mathbf{C}\tilde{\mathbf{A}}_d^j\mathbf{x}_a(0).
\end{equation}
Here, $[\mathbf{C},\tilde{\mathbf{A}},\mathbf{x}_a(0)]$ is called a realization for $\{y(j)\}$.

At last, calculate the transfer function and obtain the parameters in the Hamiltonian of Eq. (\ref{Hgen}). With $\{y(j)\}$ measured from experiments, we can obtain a realization using eigenstate realization algorithm (ERA). Through ERA, we can obtain a new realization $[\mathbf{\hat{C}},\mathbf{\hat{A}},\mathbf{\hat{x}}(0)]$, such that \begin{equation}
y(j)=\mathbf{\hat{C}}\mathbf{\hat{A}}_d^j\mathbf{\hat{x}}(0),
\end{equation}
where $\mathbf{\hat{A}}_d=\exp(\mathbf{\hat{A}} \Delta t)$. For different realizations, an invariant function called transfer function exist, which is \begin{equation}
\label{eqtr}
T(s)=\mathbf{C}(s\mathbf{I}- \mathbf{\tilde{A}})^{-1}\mathbf{x}_a(0)=\mathbf{\hat{C}}(s\mathbf{I}- \mathbf{\hat{A}})^{-1}\mathbf{\hat{x}}(0),
\end{equation}
where $s$ is called a Laplace variable  \footnote{Usually, a transfer function is denoted with $G(s)$. Here, to distinguish it from the set $G_i$, we use $T(s)$ instead.}. By comparing the coefficients of $s$, several polynomial equations can be obtained. Solving these equations leads to the identification of $a_m$.

\begin{figure} [!htb]
\includegraphics[scale=0.4]{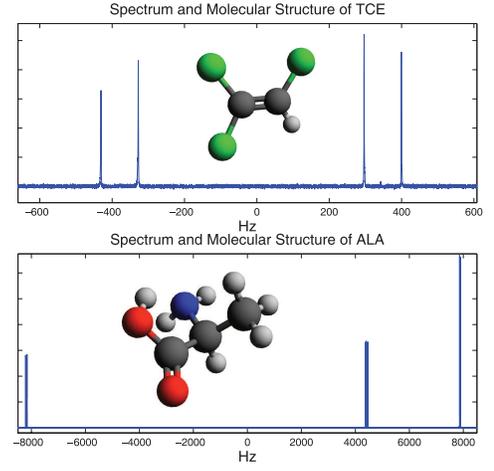}
\caption{(Coloronline) Molecule structure and spectrum for ALA and TCE. The gray balls represent $^{13}$C. Both of the spectrum are obtained with hydrogen decoupled. Note the difference of the height of the peaks of TCE spectra of one of the two carbons, which are caused by strong coupling.}
\label{molspec}
\end{figure}

\textit{Experiments setup and results.}-ZS approach was tested in a two-qubit and a three-qubit NMR systems, which were implemented with $^{13}$C-labelled trichloroethylene (TCE)  and  $^{13}$C-labelled L-alanine (ALA) as the working media, respectively. The molecular structures and the thermal spectra of ALA and TCE are shown in FIG. \ref{molspec}.

The Hamiltonian of a liquid NMR system is ($\hbar=1$) \begin{equation}
\hat{H}_{\text{NMR}}=\sum_{j=1}^{N}\pi\nu_j\sigma^{z}_{j}+\sum_{j>i=1}^{N}\frac{\pi J_{ij}}{2}{{\boldsymbol\sigma}}_{i}\cdot {{\boldsymbol \sigma}}_{j},
\end{equation}
where $2\pi\nu_{i}$ is the Larmor frequency for the $i-$th nuclei, $J_{ij}$ is the indirect spin-spin coupling constant between the $i-$th and $j-$th nuclei. For weak coupling, i.e.  $|\nu_i - \nu_j| \gg |J_{ij}|$, only the secular component of the scalar coupling is retained, i.e. $\boldsymbol{\sigma}_i \cdot \boldsymbol{\sigma}_j\approx\sigma_i^z\sigma_j^z$. The weak coupling approximation is not valid for TCE, but valid for ALA.

In this Letter, $I,X,Y$, and $Z$ are used to denote the $2\times 2$ identity matrix,$\sigma_x$, $\sigma_y$, and $\sigma_z$, respectively.  Let's write the Hamiltonian in the parametrized form following Eq. (\ref{Hgen}).
For TCE, the Hamiltonian is \begin{equation}
\hat{H}_{\text{TCE}}=a_1^{\text{T}} ZI + a_2^{\text{T}} IZ+ a_3^{\text{T}}(XX+YY+ZZ),
\end{equation}
and for ALA, the Hamiltonian is \begin{eqnarray}
\hat{H}_{\text{ALA}}&=&a_1^{\text{A}} ZII+ a_2^{\text{A}}IZI +a_3^{\text{A}}IIZ\\ \nonumber &+&a_4^{\text{A}}ZZI+a_5^{\text{A}}ZIZ+a_6^{\text{A}}IZZ.
\end{eqnarray}

Once the Hamiltonian is parametrized, we choose the observable.  For TCE with strong coupling, $O^{\text{T}}=IX$ is chosen and $\mathcal{M}^{\text{T}}=\{IX\}$, i.e., due to the strong coupling, only by the spectrum of only one qubit can the whole Hamiltonian be determined. But for ALA with weak coupling, $O^{\text{A}}=XII+IIX+XII$ has to be chosen as the observable, hence $\mathcal{M}^{\text{A}}=\{XII,IXI,IIX\}$. Following Eq. (\ref{gbar}), $\bar{G}$ can be obtained.  Once the observable is chosen, $\mathbf{\tilde{A}}$ can be decided according to Eq. (\ref{A2}).

Then we obtain the dynamics of the observable. Different from NMR quantum computing,  where experiments start from preparing a pseudo-pure state \cite{revnmr}, Hamiltonian parameter characterization starts directly states that can be easily prepared without knowing the details of the Hamiltonian,  e.g., state  $\rho(0)=\sum_j \sigma_x^j$ can be easily prepared. It should be noted is that multiple initial states instead of one initial state, which results in a complicated $\mathbf{x}_{a2}(0)$, are chosen for ALA.  For TCE, choosing $IX$ is enough. But for ALA, three initial states with density matrices $XII, IXI,$ and $IIX$ were chosen. Three initial states implies the experiment should be repeated for three times with ALA. Here, we can see that different forms of Hamiltonian lead to different experimental setups.

After the initial state $\rho(0)$ is prepared, it starts to evolve under the system Hamiltonian (and some decoherence mechanisms, which will be discussed later), hence the macroscopic magnetization rotates.  The rotation of the magnetization induces an electromagnetic wave which can be received by a coil. Due to the relaxation mechanisms, the magnitudes of the signals decays.  The voltage signal induced in the coil at time $t$ can be described as \begin{equation}
V(t)=\alpha \Tr\{F^- \rho(t)\},
\end{equation}
where $\alpha$ is a coefficient related to the static magnetic field, the species of the nuclei, and the electromagnetic properties of the receiving coil, $F^-=F_x-iF_y$ is the observable, and $\rho(t)$ is the density matrix of the system at time $t$. $F_x=\sum_j \sigma_x^{o(j)}$ and $F_y=\sum_j \sigma_y^{o(j)}$,  $o(j)$ is the number of the $j-$th nuclei being observed. $F_x$ and $F_y$ are both Hermitian operators and thus have real eigenvalues and an imginary unit $i$ was multiplied before $F_y$. Therefore, the FID has both real and imaginary parts.  The real part of the FID is\begin{equation}
V_R(t)=\alpha \Tr\{F_x \rho(t)\}.
\end{equation}
For TCE, only the second qubit is observed and for ALA, all three qubit are observed. Thus the observable for TCE is $F^{\text{T}}_x=IX$ and for ALA is $F^{\text{A}}_x=XII+IXI+IIX$, which are exactly the observables we chose for ZS approach. The state $\rho(t)$ evolves as $\rho(t)=U\rho(0)U^{\dagger}$ with $U=\exp(-i\hat{H}_{\text{NMR}}t)$. Here, the decoherence mechanism is neglected.  Detailed discussion about the influence of decoherence (described by $T_2$) will be carried out through numerical simulations. From this discussion, we can see that the FID is the measurement time traces required by ZS approach.

After the FID signals is acquired, theoretically, one can obtain a realization through ERA. However, data processing is not that simple due to the imperfections of a spectrometer. One critical thing is the dead time of the receiver, that is, the difference between the time the initial state is prepared, and the time the receiver actually records the electromagnetic signals. This could be corrected by doing  phase correction on the spectrometer. The other problem is the coefficient $\alpha$, which can be solved by scaling the FID properly. Detailed discussion about processing the experimental data before ERA is shown in the Supplementary Material.

\begin{table*}[!htb]

\begin{tabular}{|c||c|c|c|}
\hline
Parameter $a_m$ & $a_1$ &{ $a_2$} & {$a_3$} \\
\hline
Hermitian Operator $X_m$ & {$ZI$} &{ $IZ$} & {$ZZ$}\\
\hline
Result from FT $|a_i^{\text{T}}|$
 &{1180.6}&	{1081.2}	&{161.9}\\
\hline
Result from ZS approach $|a_i^{\text{E}}|$ &{1179.4}&{1082.5}&{162.6}\\
\hline
Relative Error $(\left||a_i^\text{E}|-|a_i^\text{T}|\right|)/|a_i^\text{T}|$& {$1.05\times 10^{-3}$}&{$1.25\times 10^{-3}$}&{$4.24\times 10^{-3}$}\\
\hline
\end{tabular}
\caption{Experimental Result for TCE.}
\label{TCEr}

\begin{tabular}{|c||c|c|c|c|c|c|}
\hline
Parameter $a_m$ & $a_1$ &{ $a_2$} & {$a_3$}&$a_4$ &$a_5$ &$a_6$ \\
\hline
Hermitian Operator $X_m$ & {$ZII$} &{ $IZI$} & {$IIZ$} &$ZZI$ &$ZIZ$&$IZZ$ \\
\hline
Result from FT $|a_i^{\text{T}}|$
 &25723.3&	13876.7&	24745.6&	84.8	&N/A\footnote{The value is 1.9, which can be obtained by setting the transmitter frequency to C1 and SWH to 100 Hz. With the spectrum of all three carbons, this value can not be obtained due to the low resolution caused by large SWH and the line broadening caused by $T_2$.}	&54.8
 \\
\hline
Result from ZS approach $|a_i^{\text{E}}|$ &25721.2&	13881.5	&24749.9&	84.3	&N/A\footnote{No reasonable result is available through ZS approach, which agrees well with that of FT-NMR.}&	55.7
  \\
\hline
Relative Error $(\left||a_i^\text{E}|-|a_i^\text{T}|\right|)/|a_i^\text{T}|$& {$7.98\times 10^{-6}$}&{$3.44\times 10^{-4}$}&{$1.71\times 10^{-4}$} &$5.5\times10^{-3}$ & &$1.59\times 10^{-2}$  \\
\hline
\end{tabular}
\caption{Experimental Result for ALA.}
\label{ALAr}

\end{table*}
Then ERA is performed to obtain a realization of the system on the processed FID signal. In our experiments and numerical simulations, 1,600 points are used in ERA. Compared to the number of points FT-NMR utilized, the number of the points ERA used is quite small. Once the realization is obtained, the transfer function can be calculated. By comparing the transfer functions, the parameters of the Hamiltonian are obtained. The results are shown in Table. \ref{TCEr} and Table. \ref{ALAr}.

From Table. \ref{TCEr} and \ref{ALAr}, we can see that the parameters obtained from FT and that from ZS approach agree well with each other. The absolute difference between the experimental values and theoretical values of the chemical shifts for TCE is about 1 Hz (note there is a multiplication of $\pi$ between the chemical shift and the parameters $a_j$), which is only slightly larger than the resolution of a modern FT-NMR spectrometer which is around 0.5 Hz. The relative errors for the parameters related to the chemical shifts of ALA $(a_1^{\text{A}} \sim a_3^{\text{A}})$ are smaller than that of TCE ($a_1^{\text{T}},a_2^{\text{T}}$), while at the same time, the relative errors for the parameters related to the spin-spin couplings of ALA ($a_4^{\text{A}}, a_6^{\text{A}}$) are larger than that of TCE ($a_3^{\text{T}}$). From these comparisons, we can say that the larger the value of a parameter, the more robust of the result. Despite the small relative error, the absolute differences of the chemical shifts of ALA is about $2\sim 4$ Hz, which is tiny but can be identified on modern NMR spectrometer. The extremely weak coupling can not be identified through looking at the whole spectrum (which means the observable is $XII+IXI+IIX$), both by using FT and ZS approach. This is caused by the low resolution of a spectrum with a large spectral width (SW) and the decoherence time $T_2$. SW is a parameter that decided before acquisition. Since the difference between the chemical shifts of C1 and C3 of alanine is large, to obtain the whole spectrum, the spectral width in Hz (SWH) has to be very large.  The FID resolution in Hertz FIDRES=SWH$/$TD \footnote{See Bruker manuals of Topspin for details.}. As FIDRES is proportional to SWH, large SWH means large FIDRES, hence low resolution. Combined with the line broadening  caused by decoherence, on the spectrum of FT-NMR, the weak coupling can not be identified, which means the information of $J_{13}$ is lost during the acquisition. Thus, the extreme weak coupling can not be identified.  To identify weak couplings, a small SWH (For ALA, it can be set to less than 100 Hz) can be chosen and the transmitter frequency can be set to the frequency of C1 or C3.

It is worth noting that only the absolute value of the parameters is provided in the Tables \ref{TCEr} and \ref{ALAr}. For a signal $V(t)=\cos(\omega t)$, the result of FT will give frequencies $\omega$ and $-\omega$. Therefore, to identify the sign of chemical shifts, modern NMR spectrometers use quadrature detection \cite{spindyn}. From the perspective of observables, the quadrature detection utilizes $F_y$ to assist identifying the frequencies, while during our experiments with ZS approach, only $F_x$ is observed. The signs of the spin-spin couplings can not be identified by just obtaining the 1D spectrum. To identify these signs, additional experiments such as COSY-45 or spin polarization transfer is required. From the above discussion, we can see that ZS approach gives almost the same amount of information as that of FT.

FT is not the only method utilized in data processing of NMR data. Other methods, such as maximum entropy method \cite{memnmr,menmr}, linear prediction \cite{lpnmr,lpmemnmr} are also in common use. None of these methods is as robust as FT. However, they have advantages in certain cases such as when there are only a few  data points. Our experiments show that ZS approach is also available in such case.

\begin{figure*}[!htb]
\includegraphics[scale=0.36]{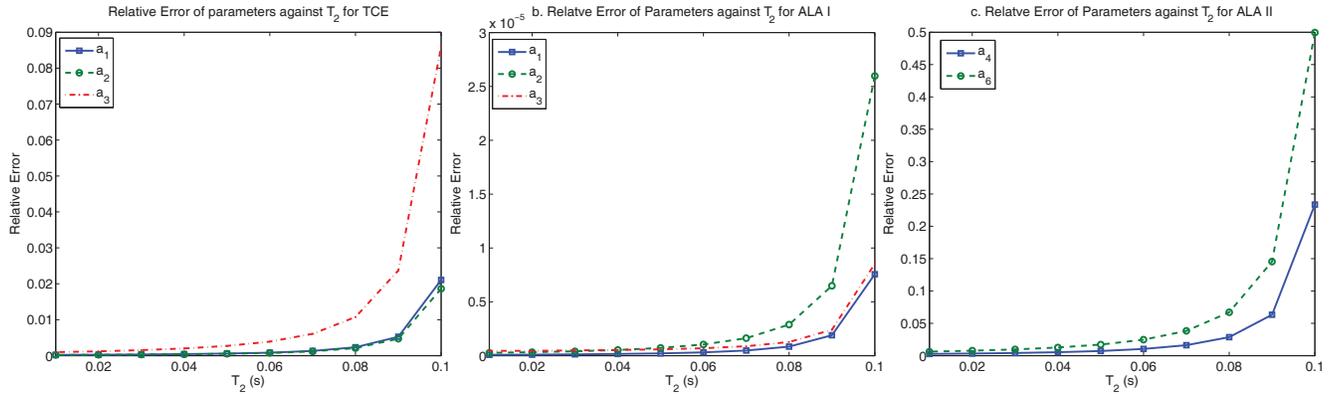}
\caption{(coloronline)Relative Error against T2. Subfigure a shows the three parameters for TCE, subfigure b shows the three parameters related to Larmor frequency of ALA, and subfigure e shows the two parameters related to the two largest coupling of ALA. }
\label{T2ben}
\end{figure*}

\textit{Decoherence.}-Then, what we want is to benchmark ZS approach with the presence of decoherence, since decoherence is a common feature for all quantum systems. One common value used to characterize decoherence is $T_2$ \footnote{In this Letter, all $T_2$ is actually $T_2^*$. $T_2^*$ consists of  $T_2$ and the inhomogeneity of the magnetic field. The inhomogeneity can be refocused by spin echo or CPMG pulses. But during the acquisition, where we obtain the FID, the inhomogeneity always exist and can not be eliminated. Thus all the $T_2$'s decribed in this Letter is $T_2^*$, which can be straightly obtained by fitting the spectra.}.  Numerical simulations are performed to benchmark the influence of $T_2$. Using the $T_2$ model presented in Ref. \cite{Ernst}, the output FID reads\begin{equation}
V_d(t)=\alpha \sum_{rs} F_{rs}^- \rho_{sr}(0) e^{\left(i\omega_{rs}-\lambda_{rs}\right)t},
\end{equation}
where  $F_{rs}^-$ $\left(\rho_{sr}(0)\right)$ denotes the $r$- and $s$-th ($s$- and $r$-th) entry of the corresponding matrix, $\omega_{rs}$ denote the frequency between energy level $r$ ans $s$ and $\lambda_{rs}$ is the relaxation rate, i.e., $\omega_{rs}=\langle r|\hat{H}_{\text{NMR}}|r\rangle-\langle s|\hat{H}_{\text{NMR}}|s\rangle$ and $\lambda_{rs}=1/T_2^{rs}$. For simplicity and without loss of generosity, let all the spins relax at the same rate, so $T_2^{rs}=T_2$ and $V_d(t)=V(t)\exp\{-t/T_2\}$. Our simulations show that, the magnitudes of the absolute value of the parameters in Hamiltonian decide the sensitivity to $T_2$. The parameters with smaller values are more sensitive to that with larger values, as shown in FIG. \ref{T2ben}.

In our simulations, the $T_2$'s are chosen to be from 0.01 s to 0.1 s, which are shorter than $T_2$ of the systems. The couplings of ALA's are much smaller than that of TCE's, which makes that the errors brought by $T_2$ to the couplings of ALA's are much larger than that of TCE's, while on the contrary, the value of the chemical shifts of ALA's are much larger than that of TCE's, so the relative errors for the chemical shifts are smaller for ALA than that for TCE. It is reasonable to find out that the longer $T_2$ is, the smaller the relative errors are.  Here, the coupling $J_{13}$ is not plotted, since for short $T_2$, it can not be identified. In classical FT-NMR, $T_2$ broadens the peak width by $1/(\pi T_2)$, which is about 3 Hz when $T_2=0.1$ s. The extreme weak coupling is about 1.8 Hz, thus the information of $J_{13}$ is lost during the acquisition due to short $T_2$.  According to our simulation, if the $T_2$ time is greater than 1.5 s, $J_{13}$ can be obtained with relative error smaller than $0.02$ using ZS approach.

\textit{Conclusion.}-In summary, we realized ZS approach using NMR quantum information processor with different work media. We showed the difference in choosing initial states and observables with different forms of couplings.  Our experiment shows that ZS approach simplifies when facing the strong coupling systems and thus this approach can be used to assist identifying the Hamiltonian systems with strong couplings, such as solid-state and liquid crystal NMR systems, the Hamiltonian of which is difficult to identify. We simultated the influence of $T_2$ on the result of ZS approach, and show the influence of $T_2$ on the result.

\begin{acknowledgments}
The authors would like to thank J. Zhang and M. Sarovar for helpful discussions. This work is supported by the National Natural Science Foundation
of China under Grants No.11474181, the National Basic Research Program
of China under Grant No. 2011CB9216002. We thank IQC, University of Waterloo, for providing the software package for NMR experiment simulation.
\end{acknowledgments}

\end{document}